\documentstyle[12pt]{article}
\def\np    { Nucl. Phys. }

\def\pl    { Phys. Lett. }

\def\beqa{\begin{eqnarray}}
\def\eeqa{\end{eqnarray}}
\def\parn              {  \par\noindent }
\def\parbigskip        {  \par\bigskip  }

\def\parbigskipn        {  \par\bigskip\noindent  }

\def\papertitlepage{\baselineskip 3.5ex \thispagestyle{empty}}
\def\Title#1{\vspace{1.5cm}\begin{center}
 {
\bf #1} \end{center} 
\vspace{1cm}}
\def\Authors#1{\begin{center} {
\it #1} \end{center}}
\def\Abstract{\vspace{0.3cm}\begin{center} {
\bf Abstract} 
           \end{center} \parbigskip}
\def\ICRRnumber#1#2#3{\hfill \begin{minipage}{3cm} #1
              \parn #2 \parn #3 \end{minipage}}
\begin{document}
\papertitlepage
\vspace*{-1 cm}
\ICRRnumber{ }{December 1999}{ }
\Title{A 10-form Gauge Potential and an M-9-brane Wess-Zumino Action \\
\vskip 1.5ex
in Massive 11D Theory} 
\Authors{{\sc\  Takeshi Sato
\footnote{tsato@icrr.u-tokyo.ac.jp}} \\
 \vskip 3ex
 Institute for Cosmic Ray Research, University of Tokyo, \\
3-2-1 Midori-cho, 
Tanashi, Tokyo 188-8502 Japan \\
}
\Abstract

We discuss some properties of 
an M-9-brane
in ``massive 11D theory''
proposed by Bergshoeff, Lozano and Ortin.
A 10-form gauge potential is consistently introduced 
into the massive 11D supergravity, and
an M-9-brane Wess-Zumino action 
is constructed as that of a gauged $\sigma$-model.
Using duality relations 
is crucial in deriving the action,
which we learn from 
the study of a 9-form potential in 10D massive IIA theory.
A target space solution of an M-9-brane with 
a non-vanishing 10-form gauge field is also obtained,
whose source is shown to be the M-9-brane effective action.

\newpage


\section{Introduction}

M-theory is conjectured to be 
the strong coupling limit of the 10 dimensional (10D)
type IIA string theory, and 
the 10D type IIA string theory
is to be 
the $S^{1}$ compactified  
M-theory in the vanishing limit of the radius\cite{wit1}. 
So, based on this conjecture,
the 10D type IIA theory and all of its constituents 
should have their 11-dimensional (11D) origins\cite{tow2}\cite{wit1}.
As is well known, 
however, the 11D origin of 10D massive IIA theory
has not been understood yet.
(The 10D massive IIA theory refers to 
the 10D IIA string theory with a nonzero
RR 10-form field strength\cite{pol2} 
and its field theory limit,  
10D massive IIA 
supergravity\cite{rom1}\cite{berg3}.)
This is due to
the fact that the massive IIA supergravity 
has a cosmological term composed of a mass parameter $m$.
Since there is the no-go theorem that  
11D supergravity forbids
a cosmological term\cite{des1},
the term (and hence the mass) 
cannot be derived from the 11D supergravity
via dimensional reduction. 
The mass is the dual of 
the field strength of a RR 9-form potential
to which a D-8-brane couples.
So, 
the 11D origin of the D-8-brane is also unclear, 
although it is conjectured to be an M-9-brane
from studies of the M-theory 
superalgebra\cite{hullalg}\cite{towalg} (see also ref.\cite{berg3}).
Some attempts have been 
made\cite{hlw1}\cite{berg4}\cite{loz1}\cite{hull3},
and one of them is
``massive 11-dimensional (11D) theory'' 
proposed by Bergshoeff, Lozano, and Ortin\cite{berg4}\cite{loz1}.
We investigate this massive theory in this paper.

First, we give a brief review of the massive 11D theory.
In this theory, target-space fields 
are required to have a Killing isometry
(whose direction is parameterized by $z$).
Its 
field theory, 
called ``massive 11D supergravity''\cite{berg4},
gives the bosonic part of the 10D massive IIA one
upon dimensional reduction in $z$.
Though this is a bosonic theory,
it is called ``supergravity''
since it reproduces the bosonic part of the 
ordinary 11D supergravity
in the massless limit 
(and if the dependence of the fields on $z$
is restored).
Target-space solutions 
of M-9-branes 
are also 
obtained\cite{bergm9}, 
which give D-8-brane solutions\cite{pol1}\cite{berg3}
when dimensionally reduced over $z$.
In addition, worldvolume actions of 
M-branes in this supergravity background have also been
examined.
In the cases of an M-wave, an M-2-brane, 
an M-5-brane and an M-Kaluza-Klein monopole,
full worldvolume actions 
are obtained as those of gauged sigma models\cite{berg4}\cite{loz1}.
These actions also give 
those of
the 10D massive IIA branes 
suggested by 
the 10D IIA superalgebra
upon direct or double dimensional reduction.\footnote{ 
Furthermore, 
actions of
branes {\it not} predicted 
by the IIA superalgebra
are also obtained\cite{eyras1}\cite{exotic}.}
In the case of an M-9-brane, however, 
{\it a Wess-Zumino (WZ) action 
has not been constructed yet},
although a kinetic term of the action has been constructed 
\cite{bergm9}\cite{eyras1}.\footnote{
It is considered that an M-9-brane 
cannot be singled out in 11 uncompactified dimensions,
but it can be singled out
when there is one compact dimension\cite{sptfilling}.
Based on the discussion on how the M-9-brane tension scales
with the radius of the 11th compact dimension\cite{bergm9},
the action in this approach describes an M-9-brane
wrapped around the compact isometry direction.}

The main purpose of this paper is
to construct the WZ action of an M-9-brane 
in a massive 11D supergravity background,
since WZ actions of branes also play important roles
in investigating properties of branes and 
dualities (e.g.  
see the recent paper \cite{lozbrantiba}). 
For this purpose,
a 10-form gauge potential is needed\cite{hullalg}
since a p-brane 
naturally couples to the (p+1)-form potential.
So, specifically,
we introduce the 10-form potential into the massive 11D
supergravity and
construct the M-9-brane WZ action by using the 10-form.

In introducing a 10-form potential, 
we follow the case of
a 9-form gauge potential
in the 10D massive IIA supergravity.
There are two methods to introduce the 9-form:
The first one is 
to promote the mass parameter $m$ to a scalar field M(x) 
and to introduce a 9-form potential as a Lagrange multiplier for the
constraint
$d M(x)=0$\cite{berg3}. 
(We denote this 9-form as $A^{(9)}$.)
The second one is superspace formulation where
the basis of the field variables of RR (p+1)-form is inspired 
by the coupling to D-p-branes\cite{bergmassiveT}\cite{DWZ}.\footnote{
The requirement of $\kappa$-symmetry of the
D-brane actions is shown to imply 
the field equations of massive IIA
supergravity\cite{superDbr}\cite{topomassive}.}
In this case the gauge transformations and the field strengths of 
dual RR gauge fields are suggested
on the basis that the RR gauge fields
can be dealt with in a geometrically uniform way,
and their consistency is checked by discussing  
T-dualities\cite{DWZ}. 
The field strength of the 9-form potential
is defined as a dual of the mass parameter $m$.
(We denote this 9-form as $C^{(9)}$.)
In this paper we choose the first one 
to introduce a 10-form potential
into the 11D theory.
The reason for this choice 
is as follows:
In the first method the massive gauge transformation of 
the 10-form potential is automatically determined.
On the other hand, if one applies the second  method
to the 11D case,
one have to construct by hand a 
massive gauge transformation
and a field strength of the 10-form consistently,
which seems difficult
since there is no geometric uniformity 
or T-duality symmetry in the massive 11D case.

However,
{\it an gauge invariant
M-9-brane WZ action cannot be constructed 
by using the 10-form 
at least straightforwardly}.
In fact this is also the case with
the D-8-brane WZ action $S^{WZ}_{D8}$ and the 9-form $A^{(9)}$
in 10D IIA theory,\footnote{Detailed discussions are given in
section 2.}
where the 9-form used to construct $S^{WZ}_{D8}$ 
is $C^{(9)}$. 
(That is,
the field redefinition relating 
the two 9-forms has not been found.) 
So, first of all, we 
explore the way to construct $S^{WZ}_{D8}$
in terms of $A^{(9)}$.
As a result, we show that
{\it 
a gauge invariant D-8-brane WZ action
can be constructed in terms of $A^{(9)}$ if 
duality relations 
are appropriately used}
(to be concrete, for rewriting 
the expression of the massive gauge transformation of $A^{(9)}$).
So, based on this lesson, 
we repeat the same procedures in the massive 11D theory:
We discuss
field strengths of 
gauge fields and their duals 
and assume their duality relations, 
based on their transformations properties 
and their relations to 10D IIA fields. 
Then, we use the relations appropriately to rewrite
the massive gauge transformation of 
$\hat{A}^{(10)}$. 
Finally, we construct a gauge invariant M-9-brane WZ action
which gives the D-8-brane WZ action on dimensional reduction
along z. 

Moreover, this paper has another purpose:
to reconstruct 
the target-space solution of a single M-9-brane
given in ref.\cite{bergm9}.
Two points are improved:
The first one is that
we solve
the equations of motion 
(of the massive 11D supergravity)
{\it with the source terms}, 
which comes from the obtained WZ action 
as well as the kinetic action of an M-9-brane.
The source terms have not been taken into account
in ref.\cite{bergm9}, but
they must be considered
since an M-9-brane can be regarded roughly as
an electric object in terms of the 10-form gauge field.
The second one is that
we construct an M-9-brane solution 
{\it with a nontrivial configuration of
10-form potential}.
The solutions in ref.\cite{bergm9}
are obtained as solutions of a {\it pure} gravity 
(i.e. only the metric field is nontrivial).
However,
this 
seems unnatural 
because usual p-brane solutions are obtained as those 
with nontrivial (p+1)-form gauge potentials.
To be concrete,
we make a certain ansatz, 
including the one done in ref.\cite{bergm9},
and solve the equations of motion
with the source terms. 

The organization of this paper is as follows: 
In section 2 we discuss the case of the 9-form in
10D massive IIA supergravity.
We first give a review of it, 
and then exhibit the problem 
and its resolution stated above explicitly.
In section 3, 
after a short review of the massive 11D supergravity,
we introduce a 10-form gauge potential. 
Then, we discuss duality relations,
use them appropriately
and construct an M-9-brane WZ action. 
In section 4 
we solve the equations of motion with source terms
and present an M-9-brane solution
with a nontrivial 10-form potential.
In section 5 we give short summary and discussion.
In the appendix, we give the relations between the 11D and the 
10D fields.
\section{A 9-form potential in
10D massive IIA theory}
In this section we begin with a brief review of the 10D massive IIA 
supergravity\cite{rom1}\cite{berg3}.
It has the same field content as the massless 
one: $\{ g_{\mu\nu}, B_{\mu\nu}, \phi, 
C^{(1)}_{\mu}, C^{(3)}_{\mu\nu\rho} 
\}$.\footnote{ 
We use mostly minus metric for 
both target-spaces and worldvolumes.
Target-space fields with hats are 11-dimensional,
and those with no hats are 10 dimensional.}
The (infinitesimal) massive gauge transformations 
are defined as
\beqa
\delta B_{\mu\nu} = 2\partial_{[\mu}\lambda_{\nu]},\ 
\delta C^{(1)}_{\mu}=-m\lambda_{\mu},\ 
\delta C^{(3)}_{\mu\nu\rho}=-3mB_{[\mu\nu}\lambda_{\rho]}
\label{10dmgt}
\eeqa
where $\lambda_{\mu}$ is a 1-form gauge parameter
and m is a constant mass parameter.
The gauge invariant field strengths of them are
\beqa 
H^{(3)}_{\mu\nu\rho}=3\partial_{[\mu}B_{\nu\rho]},& & 
G^{(2)}_{\mu\nu} = 2\partial_{[\mu}C^{(1)}_{\nu]}
+m B_{\mu\nu},\nonumber\\
G^{(4)}_{\mu\nu\rho\sigma}&=& 
4\partial_{[\mu}C^{(3)}_{\nu\rho\sigma]}
-12\partial_{[\mu}B_{\nu\rho}C^{(1)}_{\sigma]}
+3mB_{[\mu\nu}B_{\rho\sigma]}\label{10dfs1}.
\eeqa
The bosonic 
action of the massive IIA supergravity is \footnote{As discussed
in ref.\cite{bergm9},
the sign of the cosmological constant is determined by T-duality.
}
\beqa
S_{0}
&=&\frac{1}{16\pi G_{N}^{(10)}}\int d^{10}x
[\sqrt{|g|} \{ e^{-2\phi}(R-4(\partial \phi)^{2}
+\frac{1}{2\cdot 3!}(H^{(3)})^{2})\nonumber\\
& &-\frac{1}{4}(G^{(2)})^{2}
-\frac{1}{2 \cdot 4!}(G^{(4)})^{2}+\frac{1}{2} m^{2}\}\nonumber\\
& &+\frac{1}{144}\epsilon^{\mu_{1}\cdots\mu_{10}} 
\{ \partial C^{(3)}\partial C^{(3)}B
+\frac{1}{2}m\partial C^{(3)}(B)^{3}+\frac{9}{80}m^{2}(B)^{5} 
\}_{\mu_{1}\cdots\mu_{10}}]
\label{iiaaction}
\eeqa 
where $\epsilon$ is the totally antisymmetric symbol
($\epsilon^{012\cdots 9}=1$). 
The bosonic action of the 10D massless IIA
supergravity can be found by taking the limit $m\to 0$.

One way to introduce a 9-form gauge potential
is to promote
the mass parameter $m$ to a scalar field $M(x)$,
and to add 
the term 
\beqa
\Delta S =\frac{1}{16\pi G_{N}^{(10)}}\int d^{10}x 
\frac{1}{10!}\epsilon^{\mu_{1}\cdots\mu_{10}}M(x)
10\partial_{[\mu_{1}} A^{(9)}_{\mu_{2}\cdots\mu_{10}]}  
\eeqa
to the action $S_{0}$\cite{berg3}. 
Then, the field equation of $A^{(9)}$
implies that the scalar field $M(x)$ is a constant $m$,\footnote{
Strictly speaking, $M(x)$ is piecewise constant.
We discuss this point in section 4.}
So, eliminating $A^{(9)}$ 
leads 
to the field equations of the original massive IIA supergravity.
After the above procedure, the $M(x)$ field equation is
\beqa
-M&=&\frac{\epsilon^{\mu_{1}\cdots\mu_{10}}}
{\sqrt{|g|}}\{ \frac{10}{10!}\partial A^{(9)}
+\frac{1}{288}\partial C^{(3)} B^{3} +M\frac{9}{144\cdot
40}B^{5}\}_{\mu_{1}\cdots\mu_{10}}\nonumber\\
& &-\frac{1}{2}G^{(2)\mu\nu}B_{\mu\nu}
-\frac{1}{8}G^{(4)\mu\nu\rho\sigma}B_{\mu\nu}B_{\rho\sigma},
\label{10dmass1}
\eeqa
implying that 
the 10-form field strength $F^{(10)}=10 \partial A^{(9)}$
is regarded as the variable canonically conjugate to $M$. 

At this moment, the original action $S_{0}$
is no longer invariant under the 
transformations (\ref{10dmgt}). 
Instead, 
$\delta S_{0}$ is proportional to 
$\partial M$, 
as
\beqa 
\delta S_{0}&=& \frac{1}{16\pi G_{N}^{(10)}}
\int d^{10} x \sqrt{|g|}[
\partial_{\mu} M \{ G^{(2)\mu\nu}\lambda_{\nu}
+\frac{1}{2}G^{(4)\mu\nu\rho\sigma}B_{\nu\rho}\lambda_{\sigma}\}
\nonumber\\
& &-\frac{\epsilon^{\mu_{1}\cdots\mu_{10}}}{\sqrt{|g|}}
\partial_{\mu_{1}} M
\{\frac{1}{48}\partial C^{(3)}B^{2}\lambda+\frac{M}{192}B^{4}\lambda
\}_{\mu_{2}\cdots\mu_{10}}].
\eeqa
This variation can be cancelled (up to total derivative) by defining such 
a variation of the 9-form gauge potential $A^{(9)}$ as
\beqa
\delta A^{(9)}_{\mu_{1}\cdots\mu_{9}}
&=&
\sqrt{|g|}
\epsilon_{\mu_{1}\cdots\mu_{9}\mu}\{ G^{(2)\mu\nu}\lambda_{\nu}
+\frac{1}{2}G^{(4)\mu\nu\rho\sigma}B_{\nu\rho}\lambda_{\sigma}\}
\nonumber\\
& &-
9!\{\frac{1}{48}\partial C^{(3)}B^{2}\lambda
+\frac{M}{192}B^{4}\lambda\}_{[\mu_{1}\cdots\mu_{9}]}.
\label{dela9initial}
\eeqa
So, it is concluded that the 9-form potential is introduced
consistently.\footnote{
After the procedure,
the field strengths of RR gauge fields are also 
not invariant under the massive gauge transformations
due to the promotion of $m$ to $M(x)$.
However, this causes no problem because
one can replace at any time $M(x)$ for a constant $m$
by solving the $A^{(9)}$ field equation.}
We note that the first two terms of (\ref{dela9initial})
cannot be expressed as exterior products of forms.

Now, we discuss
the construction of the D-8-brane WZ action $S_{D8}^{WZ}$.
What we want to do is 
to construct
$S_{D8}^{WZ}$ by using the 9-form $A^{(9)}$.
However, this cannot be achieved at least straightforwardly.
To be concrete,
since a D-8-brane couples to a 9-form potential,
$S_{D8}^{WZ}$ contains the term
\beqa
S_{D8}^{WZ}|_{{\rm 9-form\  part}}
=\frac{T}{9!}\int d^{9}\xi\  \epsilon^{i_{1}\cdots i_{9}}
\partial_{i_{1}}X^{\mu_{1}}\cdots \partial_{i_{9}}X^{\mu_{9}} 
A^{(9)}_{\mu_{1}\cdots \mu_{9}}\label{d8brwz}
\eeqa
where $\xi_{i}$ ($i=0,..,8$) are worldvolume coordinates
of the brane and $X^{\mu} \ \ (\mu=0,..,9)$ are embedding coordinates.
Suppose we consider the massive gauge transformation of (\ref{d8brwz}).
Then, we can see that 
the contribution of the first two terms of (\ref{dela9initial})
to the variation of $S_{D8}^{WZ}|_{{\rm 9-form\  part}}$
cannot be cancelled
even if any other terms are added to 
$S_{D8}^{WZ}|_{{\rm 9-form\  part}}$.
This implies that one cannot keep the massive gauge
symmetry of the brane action if $A^{(9)}$ is used straightforwardly.

Our idea to resolve this problem is very simple:
if one rewrite the first two terms of (\ref{dela9initial})
by using 
the dual fields of $C^{(1)}$ and $C^{(3)}$
through duality relations,
the two terms can be expressed 
as a sum of exterior products of forms.
So, it is expected that
a gauge invariant WZ action can be constructed.

This idea is a success,
which we show in the following:
The dual fields of $C^{(3)}$ and $C^{(1)}$
is the 5-form $C^{(5)}$ and
the 7-form $C^{(7)}$, respectively.
We use the duals defined in ref.\cite{DWZ},
whose field strengths and massive gauge transformations
are given respectively as
\[
\left. 
\begin{array}{cc}
\left\{
\begin{array}{cc}
G^{(6)}_{\mu_{1}\cdots\mu_{6}}
=&(6\partial C^{(5)}-60\partial B C^{(3)} +15 M B^{3}
)_{[\mu_{1}\cdots\mu_{6}]},\\  
G^{(8)}_{\mu_{1}\cdots\mu_{8}}=&(
8\partial C^{(7)}-168\partial B C^{(5)}+105 M B^{4}
)_{[\mu_{1}\cdots\mu_{8}]},
\end{array}
\right. 
&
\left\{
\begin{array}{cc}
\delta C^{(5)}_{\mu_{1}\cdots\mu_{5}}=& -15 (B^{2}\lambda
)_{[\mu_{1}\cdots\mu_{5}]},\\ 
\delta C^{(7)}_{\mu_{1}\cdots\mu_{7}}=& -105 (B^{3}\lambda
)_{[\mu_{1}\cdots\mu_{7}]}.
\end{array}
\right.
\end{array}
\right.
\]
The duality relations of them (in our notation) are\cite{DWZ}
\beqa
G^{(2)\mu_{1}\mu_{2}}=-\frac{\epsilon^{\mu_{1}\cdots
\mu_{10}}}{8!\sqrt{|g|}}G^{(8)}_{\mu_{3}\cdots
\mu_{10}},\ \ \ 
G^{(4)\mu_{1}\cdots\mu_{4}}=\frac{\epsilon^{\mu_{1}\cdots
\mu_{10}}}{6!\sqrt{|g|}}G^{(6)}_{\mu_{5}\cdots
\mu_{10}}.\label{10ddualityrel}
\eeqa
Substituting these relations for the first two terms 
of (\ref{dela9initial}) leads to the rewritten
expression 
\beqa
\delta A^{(9)}_{\mu_{1}\cdots\mu_{9}}=-9![
\frac{1}{7!}\partial C^{(7)}\lambda
-\frac{1}{2\cdot 5!}\partial(C^{(5)}B)\lambda
+\frac{1}{2^{3}\cdot 3!}\partial(C^{(3)}B^{2})\lambda
-\frac{M}{2^{4}\cdot 4!}B^{4}\lambda]_{[\mu_{1}\cdots\mu_{9}]}.\ \ 
\label{dela9initial2}
\eeqa
Using this expression,
we can obtain 
the WZ action invariant under the massive gauge transformations.
This means that via a field redefinition, 
$A^{(9)}$ is related
to the 9-form potential $C^{(9)}$ 
which is usually used for constructing $S_{D8}^{WZ}$.
So, instead of giving the action in terms of $A^{(9)}$,
we present the redefinition relation.
The field strength of $C^{(9)}$ is defined in this case as\cite{DWZ}
\beqa
-M=\ast G^{(10)}
=\frac{\epsilon^{\mu_{1}\cdots\mu_{10}}}{10!\sqrt{|g|}}
10 [ \partial C^{(9)}-36\partial B C^{(7)}
+\frac{189M}{2}B^{5} ]_{\mu_{1}\cdots\mu_{10}}.\label{10dmass2}
\eeqa
Substituting (\ref{10ddualityrel}) for
the last two terms of (\ref{10dmass1}) and
comparing it with (\ref{10dmass2}),
we can determine 
the relation between the two
9-forms up to total derivative as
\beqa
C^{(9)}_{\mu_{1}\cdots\mu_{9}}
=\{ A^{(9)}+9!(\frac{1}{2\cdot 7!}C^{(7)}B
-\frac{1}{2^{3}\cdot 5!}C^{(5)}B^{2}
+\frac{1}{2^{3}(3!)^{2}}C^{(3)}B^{3}) \}_{[\mu_{1}\cdots\mu_{9}]}
\label{10dredef}.
\eeqa
This redefinition relation is consistent with
the massive gauge transformations of the two,
(i.e. (\ref{dela9initial2}) and 
$ \delta C^{(9)}_{\mu_{1}\cdots\mu_{9}}=
-945M(B^{4}\lambda)_{[\mu_{1}\cdots\mu_{9}]}$).
So, $S_{D8}^{WZ}$ can be constructed via $A^{(9)}$.
Thus, we conclude that {\it using duality relations are crucial to
derive the the WZ action in terms of the 9-form $A^{(9)}$}.

\section{A 10-form gauge potential 
and an M-9-brane Wess-Zumino action in massive 11D supergravity}
\setcounter{footnote}{0} 
In this section we first review the massive 11D
supergravity\cite{berg4}.
The bosonic field content of the supergravity is the same 
as that of the usual (massless) 11D
supergravity:
the metric $\hat{g}_{\mu\nu}$  and a 3-form gauge potential
$\hat{C}_{\mu\nu\rho}$. 
In this theory these fields are required to have a Killing isometry,
i.e., 
${\cal L}_{\hat{k}}\hat{g}_{\mu\nu}
={\cal L}_{\hat{k}}\hat{C}_{\mu\nu\rho}=0$
where $ {\cal L}_{\hat{k}}$ indicates a Lie derivative 
with respect to a Killing vector field $\hat{k}^{\mu}$.
(We take the coordinates so that $\hat{k}^{\mu}=\delta^{\mu z}$.)
The infinitesimal gauge transformations of the fields are
defined as 
\beqa
\delta\hat{g}_{\mu\nu}=-m[\hat{\lambda}_{\mu}(i_{\hat{k}}\hat{g})_{\nu}
+\hat{\lambda}_{\nu}(i_{\hat{k}}\hat{g})_{\mu}], 
\ \ \delta\hat{C}_{\mu\nu\rho}=3
\partial_{[\mu}\hat{\chi}_{\nu\rho]}
-3m\hat{\lambda}_{[\mu}(i_{\hat{k}}\hat{C})_{\nu\rho]}
\label{massivegt0}
\eeqa
where 
$(i_{\hat{k}}T^{(r)}_{\mu_{1}\cdots \mu_{r-1}})\equiv\hat{k}^{\mu}
T^{(r)}_{\mu_{1}\cdots \mu_{r-1}\mu}$ for a field $T^{(r)}$.
$\hat{\chi}$ is the infinitesimal 2-form gauge parameter,
and $\hat{\lambda}$ is defined as $\hat{\lambda}_{\mu}
\equiv (i_{\hat{k}}\hat{\chi})_{\mu}$.\footnote{In this paper
we change the notation
of ref.\cite{berg4}
such that $m\to 2m$ and $\hat{\lambda}\to -\frac{1}{2}\hat{\lambda}$.
}
Then, a connection for the massive gauge transformations  
should be considered.
The new total connection takes the form 
$\hat{\Omega}_{a}^{\ bc}=\hat{\omega}_{a}^{\ bc}
+\hat{K}_{a}^{\ bc}$\footnote{
We use $a,b,\cdots$ for local Lorentz indices.}
where $\hat{\omega}_{a}^{bc}$ is a usual spin connection
and
$\hat{K}$ is given by
\beqa
\hat{K}_{a}^{\ bc}=\frac{m}{2}
[\hat{k}_{a}(i_{\hat{k}}\hat{C})^{bc}
+\hat{k}_{b}(i_{\hat{k}}\hat{C})^{ac}
-\hat{k}_{c}(i_{\hat{k}}\hat{C})^{ab}].
\eeqa
The 4-form field strength $\hat{G}^{(4)}$ of $ \hat{C}$
is defined as 
\beqa
\hat{G}^{(4)}_{\mu\nu\rho\sigma}=4D_{[\mu}\hat{C}_{\nu\rho\sigma]}
\equiv 4\partial_{[\mu}\hat{C}_{\nu\rho\sigma]}
+3m(i_{\hat{k}}\hat{C})_{[\mu\nu}(i_{\hat{k}}\hat{C})_{\rho\sigma]}
\eeqa
where $D_{\mu}$ denotes the covariant derivative.
Then, $\hat{G}^{(4)}$ transforms covariantly as
\beqa
\delta \hat{G}^{(4)}_{\mu\nu\rho\sigma}=4m\hat{\lambda}_{[\mu}
(i_{\hat{k}}G^{(4)})_{\nu\rho\sigma]},
\eeqa
which implies that $\delta (\hat{G}^{(4)})^{2}=0$.

The action of the massive 11D supergravity is 
\beqa
\hat{S}_{0}&=&\frac{1}{\kappa}
\int d^{11}x [\ \sqrt{|\hat{g}|}\{ \hat{R}
-\frac{1}{2\cdot 4!}(\hat{G}^{(4)})^{2}
+\frac{1}{2}m^{2}|\hat{k}|^{4} \} \nonumber\\
&+&\frac{\hat{\epsilon}^{
\mu_{1}\cdots \mu_{11}}}{(144)^{2}}
\{2^{4}\partial\hat{C}\partial\hat{C}\hat{C}
+18m\partial\hat{C}\hat{C}(i_{\hat{k}}\hat{C})^{2}
+\frac{3^{3}}{5}m^{2}\hat{C}(i_{\hat{k}}
\hat{C})^{4}\}_{
\mu_{1}\cdots \mu_{11}}]\ \ \ \ 
\label{11daction0}
\eeqa
where $\kappa =16\pi G_{N}^{(11)}$ and $|\hat{k}|
=\sqrt{-\hat{k}^{\mu}\hat{k}^{\nu}\hat{g}_{\mu\nu}}$.
This action is invariant (up to total derivative)
under (\ref{massivegt0}). 
The dimensional reduction of 
the action 
along $z$
is shown to give the bosonic part of 10D massive IIA 
supergravity.\footnote{
By using the (generalized) Palatini's identity
given in ref.\cite{berg4},
the action (\ref{11daction0}) is 
rewritten as
\beqa
\hat{S}_{0}&=&\frac{1}{\kappa}
\int d^{11}x [\sqrt{|\hat{g}|}
\{- \hat{\Omega}_{b}^{\ ba}\hat{\Omega}_{c\ a}^{\ c}
-\hat{\Omega}_{a}^{\ bc}\hat{\Omega}_{bc}^{\ \ a}
-\frac{1}{2\cdot 4!}(\hat{G}^{(4)})^{2}
+\frac{1}{2}m^{2}|\hat{k}^{2}|^{2} \} \}\nonumber\\
&+&\frac{1}{144}\hat{\epsilon}^{\mu_{1}\cdots\mu_{10} z}
\{ \partial\hat{C}\partial\hat{C}(i_{\hat{k}}\hat{C})
+\frac{m}{2}\partial\hat{C}(i_{\hat{k}}\hat{C})^{3}
+\frac{9m^{2}}{80}(i_{\hat{k}}\hat{C})^{5}
\}_{\mu_{1}\cdots\mu_{10} z}
+({\rm surface\ \  terms})]. \nonumber
\eeqa
In fact, the 10D action (\ref{iiaaction}) is obtained from this action
only if the surface terms are omitted.
Omitting them, we use this action as
a ``starting'' action, in order to 
make the correspondence of the 11D theory with the 10D one.
}
(See the appendix for the relation between the 11D and 10D fields.)

Now, let us introduce a 10-form gauge potential $\hat{A}^{(10)}$.
Following the case of the 9-form potential in 10D IIA theory,
we promote
the mass parameter $m$ to a scalar field $\hat{M}(x)$,
and add the term 
\beqa
\Delta S =\frac{1}{\kappa}\int d^{11}x 
\frac{1}{11!}\hat{\epsilon}^{\mu_{1}\cdots\mu_{11}}\hat{M}(x)
11\partial_{[\mu_{1}} \hat{A}^{(10)}_{\mu_{2}\cdots\mu_{11}]}.  
\eeqa
We note that $\hat{A}^{(10)}$ also satisfies $ {\cal
L}_{\hat{k}}\hat{A}^{(10)}=0$, which means that
$\hat{A}^{(10)}$ with no $z$ index does not appear
in this theory.
Then, the action is invariant under (\ref{massivegt0})
if the massive gauge transformation of $\hat{A}^{(10)}$
is defined as
\beqa
\delta (i_{\hat{k}}\hat{A}^{(10)})_{\mu_{1}\cdots\mu_{9}}
&=&-\sqrt{|\hat{g}|}
\hat{\epsilon}_{\mu_{1}\cdots\mu_{9}\mu z}
[ 
-\hat{g}^{\mu\mu'}\hat{g}^{\nu\nu'}
(2\partial_{[\mu'} \hat{k}_{\nu']}
-\hat{M}|\hat{k}|^{2}(i_{\hat{k}}\hat{C})_{\mu'\nu'})
\hat{\lambda}_{\nu}\nonumber\\
&+&\frac{1}{2}\hat{G}^{(4)\mu\nu\rho\sigma}
(i_{\hat{k}}\hat{C})_{\nu\rho}\hat{\lambda}_{\sigma}
]
-\frac{9!}{48}[\partial \hat{C}(i_{\hat{k}}\hat{C})^{2}\hat{\lambda}
+\frac{\hat{M}}{4}(i_{\hat{k}}\hat{C})^{4}
\hat{\lambda}]_{\mu_{1}\cdots\mu_{9}}\label{10formtr2}\\
\delta \hat{A}^{(10)}_{\mu_{1}\cdots\mu_{10}} &=&10\hat{M}
\hat{\lambda}_{[\mu_{1}}
(i_{\hat{k}}\hat{A}^{(10)})_{\mu_{2}\cdots\mu_{10}]}
{\rm \ \ \ \ when \ \  \mu_{1},\cdots,\mu_{10}\ne z }
\label{10formtr1} 
\eeqa
where (\ref{10formtr1}) 
is the expected massive gauge transformation
of a 10-form gauge field.


Next, we prepare to
rewrite the first two terms of 
the massive transformation (\ref{10formtr2}), 
in the same way as the 10D case.
The dual field of the 3-form $\hat{C}$
is the 6-form $\hat{C}^{(6)}$ 
whose massive gauge transformation, field strength and
the duality relation are\cite{berg4}\footnote{We
concentrate our discussions on the gauge 
transformations with respect to
$\hat{\chi}$ and $\hat{\lambda}$.} 
\beqa
\delta \hat{C}^{(6)}_{\mu_{1}\cdots\mu_{6}}&=&30
\partial_{[\mu_{1}}\hat{\chi}_{\mu_{2}\mu_{3}}
\hat{C}_{\mu_{4}\mu_{5}\mu_{6}]}  
+6\hat{M}\hat\lambda_{[\mu_{1}}
(i_{\hat{k}}\hat{C}^{(6)})_{\mu_{2}\cdots\mu_{6}]}\\
\hat{G}^{(7)}_{\mu_{1}\cdots\mu_{7}}&=&7[
\partial\hat{C}^{(6)}
-3\hat{M}(i_{\hat{k}}\hat{C})(i_{\hat{k}}C^{(6)})
+10\hat{C}\partial \hat{C}
+5\hat{M}C(i_{\hat{k}}\hat{C})^{2}
+\frac{\hat{M}}{7}(i_{\hat{k}}\hat{N}^{(8)})]_{\mu_{1}\cdots\mu_{7}}\ \ \ \
\ \ \ \ \\
\hat{G}^{(4)\mu_{1}\cdots\mu_{4}}&=&\frac{\epsilon^{\mu_{1}\cdots
\mu_{11}}}{7!\sqrt{|g|}}\hat{G}^{(7)}_{\mu_{5}\cdots
\mu_{11}}\label{11ddual1}.
\eeqa
$\hat{N}^{(8)}$ is 
the dual field of the Killing vector also introduced in
ref.\cite{berg4}, whose gauge transformation is suggested such as
\beqa
\delta \hat{N}^{(8)}_{\mu_{1}\cdots\mu_{8}}=\{
\frac{8!}{3\cdot 4!}\partial \hat{\chi}\hat{C}(i_{\hat{k}}\hat{C})
 +8\hat{M}\hat{\lambda}(i_{\hat{k}}\hat{N}^{(8)})\}_{\mu_{1}\cdots\mu_{8}}.
\eeqa
In this paper 
we regard 
$\hat{k}_{\mu}\equiv (i_{\hat{k}}\hat{g})_{\mu}$ as a ``vector
gauge field'', 
and consider the ``field strength'' of it,
as done for 
$(i_{\hat{k}}\hat{C})$.
Then, if we define $\hat{G}^{(2)}$
as
\beqa
\hat{G}^{(2)}_{\mu\nu}\equiv 2\partial_{[\mu}\hat{k}_{\nu]}
-\hat{M}|k|^{2}(i_{\hat{k}}\hat{C})_{\mu\nu},
\eeqa
$\hat{G}^{(2)}$ is shown to 
transform covariantly under (\ref{massivegt0}).
So, $\hat{G}^{(2)}$, in fact 
arising in the first term of (\ref{10formtr2}),
can be interpreted as the field strength of 
$\hat{k}_{\mu}$.
On the other hand,
the field strength $\hat{G}^{(9)}$ 
of the full 8-form $\hat{N}^{(8)}$
is difficult to construct.
However, in order to rewrite the first term through the duality
relation between $\hat{G}^{(9)}$ and $\hat{G}^{(2)}$,
it is sufficient to know
the field strength of $(i_{\hat{k}}\hat{N}^{(8)})$. 
This is because $\hat{G}^{(2)}$
in the first term of (\ref{10formtr1})
vanishes if
any of the indices of $\hat{G}^{(2)}$ takes $z$,
implying that one of the indices of $\hat{G}^{(9)}$ certainly 
takes $z$. 
Thus,
only the field strength of $(i_{\hat{k}}\hat{N}^{(8)})$ is needed.
It can be defined as
\beqa
(i_{\hat{k}}\hat{G}^{(9)})_{\mu_{1}\cdots\mu_{8}}&\equiv&
8\{ \partial (i_{\hat{k}}\hat{N}^{(8)})
+21(i_{\hat{k}}\hat{C}^{(6)})\partial(i_{\hat{k}}\hat{C})
\nonumber\\
&+&35C\partial(i_{\hat{k}}\hat{C})(i_{\hat{k}}\hat{C})
+35\partial C(i_{\hat{k}}\hat{C})^{2}
+\frac{105}{8}\hat{M}(i_{\hat{k}}\hat{C})^{4}
\}_{[\mu_{1}\cdots\mu_{8}]}.\label{n8fs}
\eeqa
We note that $(i_{\hat{k}}\hat{G}^{(9)})$ 
is invariant under (\ref{massivegt0}), which means that this
definition is consistent.
Then, we assume the duality relation: 
\beqa
\hat{G}^{(2)\mu_{1}\mu_{2}}=\frac{\epsilon^{\mu_{1}\cdots
\mu_{10}z}}{9!\sqrt{|g|}}
(i_{\hat{k}}\hat{G}^{(9)})_{\mu_{3}\cdots\mu_{10}}\label{11ddual2}.
\eeqa
It gives
one of the 10D IIA duality relations in (\ref{10ddualityrel})
on dimensional reduction in $z$, which means that (\ref{11ddual2}) 
is consistent. 

Since all the preparations have been done,
let us substitute the relation
(\ref{11ddual1}) and (\ref{11ddual2}) for
(\ref{10formtr2})
to have the rewritten expression
of the massive gauge transformation of $\hat{A}^{(10)}$:
\beqa
\delta (i_{\hat{k}}\hat{A}^{(10)})_{\mu_{1}\cdots\mu_{9}}
&=&-9![\ 
\frac{1}{7!}\partial(i_{\hat{k}}\hat{N}^{(8)})\hat{\lambda}
-\frac{1}{2\cdot 5!}\partial\{(i_{\hat{k}}\hat{C}^{(6)})
(i_{\hat{k}}\hat{C})\}\hat{\lambda}\nonumber\\
&+&\frac{1}{6\cdot 4!}\partial\{\hat{C}(i_{\hat{k}}\hat{C})^{2}\}
\hat{\lambda}
-\frac{\hat{M}}{2^{4}\cdot 4!}(i_{\hat{k}}\hat{C})^{4}\hat{\lambda}\ 
]_{[\mu_{1}\cdots\mu_{9}]}.
\eeqa
By using this expression,
the gauge invariant WZ action of the M-9-brane can be constructed.
Before constructing it, we give
the rewritten field equation of $\hat{M}(x)$:
\beqa
-\sqrt{|\hat{g}|}\hat{M}|\hat{k}|^{4}
&=&\frac{10}{10!}
\hat{\epsilon}^{\mu_{1}\cdots\mu_{10}z}\{
\partial_{\mu_{1}}
(i_{\hat{k}}\hat{A}^{(10)})_{\mu_{2}\cdots\mu_{10}}
-\frac{9!}{8\cdot 6!}(i_{\hat{k}}\hat{G}^{(7)})(i_{\hat{k}}\hat{C})^{2}
+\frac{9!}{2\cdot 8!}(i_{\hat{k}}\hat{G}^{(9)})(i_{\hat{k}}\hat{C})
\nonumber\\
& &+\frac{9!}{288}
\partial\hat{C}(i_{\hat{k}}C^{(3)})^{3}
+\frac{9\cdot 9!}{144\cdot 40}\hat{M}(i_{\hat{k}}C^{(3)})^{5}
\}_{\mu_{1}\cdots\mu_{10}}.\label{fs11form}
\eeqa
Since the right hand side of (\ref{fs11form}) 
is shown to be gauge invariant,
it can be interpreted as the gauge invariant field strength of 
the 10-form (multiplied by 1/10!).
Thus, we can conclude that the 10-form $\hat{A}^{(10)}$
is introduced consistently.
Moreover,
we define 
a new 10-form $\hat{C}^{(10)}$
which coincides with 10D IIA 9-form $C^{(9)}$
on dimensional reduction along z:
\beqa
(i_{\hat{k}}\hat{C}^{(10)})_{\mu_{1}\cdots\mu_{9}}
&\equiv&(i_{\hat{k}}\hat{A}^{(10)})_{\mu_{1}\cdots\mu_{9}}
+9!\ [\frac{1}{2\cdot
7!}(i_{\hat{k}}\hat{N}^{(8)})(i_{\hat{k}}\hat{C})
\nonumber\\
& &-\frac{1}{2^{3}\cdot 5!}(i_{\hat{k}}\hat{C}^{(6)})
(i_{\hat{k}}\hat{C})^{2}
+\frac{1}{2^{4}\cdot (3!)^{2}}
\hat{C}(i_{\hat{k}}\hat{C})^{3}]_{[\mu_{1}\cdots\mu_{9}]}
\nonumber\\
\hat{C}^{(10)}_{\mu_{1}\cdots\mu_{10}}&\equiv&
\hat{A}^{(10)}_{\mu_{1}\cdots\mu_{10}}.
\eeqa
Then, the gauge transformation of $\hat{C}^{(10)}$ takes the simple form:
\beqa
\delta (i_{\hat{k}}\hat{C}^{(10)})_{\mu_{1}\cdots\mu_{9}}
=-945\{-4\partial\chi(i_{\hat{k}}\hat{C})^{3}
+ \hat{M}(i_{\hat{k}}\hat{C})^{4}\hat{\lambda}\ 
\}_{[\mu_{1}\cdots\mu_{9}]}.
\eeqa 
For convenience, we use this 10-form 
to construct $S^{WZ}_{{\rm M9}}$.

Finally, we construct the M-9-brane WZ action
as that of the gauged $\sigma$-model,
in which the translation along $\hat{k}$ is 
gauged\cite{kaluzakleinm}\cite{berg4}\cite{loz1}.
In this approach the M-9-brane
wrapped around the compact isometry direction is 
described\cite{bergm9}.
So, denoting its worldvolume coordinates by $\xi^{i} \  (i=0,1,..,8)$
and their embeddings by $X^{\mu}(\xi) (\mu =0,1,..,9,z)$,
the 
worldvolume gauge
transformation is given by
\beqa
\delta_{\eta} X^{\mu}=\eta(\xi)\hat{k}^{\mu}\label{wvgaugetr}
\eeqa
where 
$\eta(\xi)$ is a scalar gauge parameter.
In order to make the brane action invariant
under the transformation,
the derivative of $X^{\mu}$ with respect to $\xi^{i}$ is 
replaced by
the covariant derivative\cite{loz1} 
\beqa
D_{i}X^{\mu}=\partial_{i}X^{\mu}
-\hat{A}_{i}\hat{k}^{\mu} 
\eeqa
with the gauge field $\hat{A}_{i}=-|\hat{k}|^{-2}
\partial_{i}X^{\nu}\hat{k}_{\nu}$.
The dimensional reduction of $D_{i}X^{\mu}$ is such that
$D_{i}X^{\mu}=
\partial_{i}X^{\mu}$ for $\mu\ne z$ and 
$D_{i}X^{z} 
=-\partial_{i}X^{\mu}C^{(1)}_{\mu}$.
($\partial_{i}X^{z}=0$
since $X^{z}$ corresponds to $z$.)
The M-9-brane action must be so constructed as
to give the D-8-brane action on dimensional reduction
along $z$.
So, considering the field relations given in the appendix,
we obtain
the M-9-brane WZ action for a constant mass background $\hat{M}(x)=m$:
\beqa
S_{M9}^{WZ}\ \ \ =\ \ \ T_{{\rm M9}}
\int d^{9}\xi \epsilon^{i_{1}\cdots i_{9}} 
[\frac{1}{9!}\widetilde{(i_{\hat{k}}\hat{C}^{(10)})}_{i_{1}\cdots i_{9}}
+\frac{1}{2\cdot 7!}
\widetilde{(i_{\hat{k}}\hat{N}^{(8)})}_{i_{1}\cdots i_{7}}
\hat{{\cal K}}^{(2)}_{i_{8}i_{9}}\ \ \ \ \ \ \ \ \ \ \ \ \ \ \ 
\ \ \ \ \ \ \  
\nonumber\\
+\frac{1}{2^{3}\cdot 5!}\widetilde{(i_{\hat{k}}\hat{C}^{(6)})}_{
i_{1}\cdots i_{5}}
(\hat{{\cal K}}^{(2)})^{2}_{{i_{6}\cdots i_{9}}} 
+\frac{1}{2\cdot (3!)^{2}}\widetilde{\hat{C}}_{i_{1} i_{2} i_{3}}
\{(\partial \hat{b})^{2} 
-\frac{1}{4}\widetilde{(i_{\hat{k}}\hat{C})}
\partial \hat{b} 
+\frac{1}{8}\widetilde{(i_{\hat{k}}\hat{C})}^{2}\}_{i_{4}\cdots i_{7}}
\hat{{\cal K}}^{(2)}_{i_{8} i_{9}}\nonumber\\
+\frac{1}{2\cdot 4!}\hat{A}_{i_{1}}
\{ (\partial \hat{b})^{3} 
+\frac{1}{2}(\partial \hat{b})^{2}\widetilde{(i_{\hat{k}}\hat{C})}
+\frac{1}{4}(\partial \hat{b})\widetilde{(i_{\hat{k}}\hat{C})}^{2}
+\frac{1}{8}\widetilde{(i_{\hat{k}}\hat{C})}^{3}
\}_{i_{2}\cdots i_{7}}
(\hat{{\cal K}}^{(2)})_{i_{8} i_{9}}\nonumber\\
+\frac{m}{5!}\hat{b}_{i_{1}}(\partial \hat{b})^{4}_{i_{2}\cdots
i_{9}}]\ \ \ \ \ \ \ \ \ \ 
\label{m9action}
\eeqa
where $\widetilde{\hat{S}}_{i_{1}\cdots i_{r}}\equiv \frac{1}{r!}
\hat{S}_{\mu_{1}\cdots\mu_{r}}
D_{i_{1}}X^{\mu_{1}}\cdots D_{i_{r}}X^{\mu_{r}}$
for a target-space field $\hat{S}_{\mu_{1}\cdots\mu_{r}}$.
$\hat{b}_{i}$ 
describes the flux 
of an M-2-brane wrapped around the isometry direction,
whose massive gauge transformation is determined by the 
requirement of the invariance of its modified field strength
$\hat{{\cal K}}^{(2)}_{ij}=2\partial_{[i}\hat{b}_{j]}-
\partial_{i}X^{\mu}\partial_{j}X^{\nu}
(i_{\hat{k}}\hat{C})_{\mu\nu}$ (i.e. $\delta\hat{b}_{i}
=\hat{\lambda}_{i}$).
We can see that this action is invariant (up to total derivative) 
under both the massive 
and the worldvolume gauge transformations (\ref{wvgaugetr}).
If one consider a worldvolume 8-form $\hat{\omega}^{(8)}$
and add the term $\int 9 d \hat{\omega}^{(8)}$ to (\ref{m9action}),
the total derivative can be compensated and the action becomes
exactly invariant.
For later use, we also present 
the kinetic part of the M-9-brane action\cite{eyras1}:
\beqa
S=-T_{{\rm M9}}\int d^{9}\xi|\hat{k}|^{3}
\sqrt{|{\rm det}(D_{i}X^{\mu}D_{j}X^{\nu}
\hat{g}^{\mu\nu}+ 
|k|^{-1}\hat{{\cal K}^{(2)}_{ij}})|}.\label{m9kinetic}
\eeqa

\section{A target-space M-9-brane solution}
In this section we obtain an M-9-brane solution 
with a nontrivial 10-form,
by solving the equations of motion 
with the source terms 
(i.e. $\delta S^{{\rm total}}\equiv
\delta\{S_{{\rm massive\  SUGRA}}
+S_{M9}\}=0$).

We set the ansatz for the target-space fields:
\beqa
ds^{2}&=&H^{\alpha}(dt^{2} -dx^{2}_{(8)})
-H^{\beta}dy^{2}-H^{\gamma}dz^{2}\nonumber\\
\hat{A}^{(10)}_{01,..,8z}&=&\hat{A}^{(10)}_{01,..,8z}(y)
\eeqa
with all the other fields vanishing, and for the worldvolume
fields: 
\beqa
X^{i}=\xi^{i}\ \  {\rm  for}\ \  i=0,1,..,8,\ \ \   
{\rm and}\ \    \hat{b}_{j}=0.
\eeqa
$H$ is a function depending 
on the single transverse direction $y$.
Then, the nontrivial
equations of motion are 
only those for $\hat{g}_{\mu\nu},\ 
\hat{A}^{(10)},\ \hat{M}$, and $X^{i}$. 

Suppose the single M-9-brane we consider lies at $y=0$.
Then, the $\hat{A}^{(10)}$ equation is\footnote{
We note that because of the existence of the isometry direction,
$\frac{\delta L(x')}{\delta L(x)}=\delta^{(10)}(x-x')$ for a field $L$,
instead of $\delta^{(11)}(x-x')$.
Moreover, the integration with respect to 
$x^{z}=z$ in the supergravity action is
performed at the beginning.}
\beqa
\partial_{y}\hat{M}=-T_{{\rm M9}}\bar{\kappa}\delta(y), 
\ \ \ 
\partial_{\mu}\hat{M}=0 {\rm \ \ \  for\ } \mu\ne y
\eeqa
where $1/\bar{\kappa}\equiv \int dx^{z}/\kappa$.
If, for simplicity, we take the symmetry between 
the region $y>0$ and $y<0$ into account,
it is proper to take 
the solution
\[
\hat{M}(x)=\left\{
\begin{array}{rl}
-\bar{m} & \quad \mbox{for $y>0$}\\ 
\bar{m} & \quad \mbox{for $y<0$}
\end{array}\right. \]
where $\bar{m}\equiv T_{{\rm M9}}\tilde{\kappa}/2$.
In other words, the mass parameter is determined by the background
M-9-brane as the above. 
Then, the Einstein equation parts of the
field equations 
of $\hat{g}_{\mu\nu}$ are the same as those before 
introducing the 10-form. 
So, we have $\alpha=-\epsilon/3, \ \beta=-10\epsilon/3-2$ and
$\gamma=5\epsilon/3$ 
for a nonzero parameter $\epsilon$, as given in ref.\cite{bergm9}.
Then, the field equations of $\hat{g}_{\mu\nu}$ are
\beqa
\frac{\delta S^{{\rm total}}}{\delta \hat{g}_{\mu\nu}}&=&
\frac{1}{2\bar{\kappa}}H^{2\epsilon/3}\eta_{\mu\nu}
[\epsilon \partial_{y}^{2}H+\frac{1}{2}H^{-1}
\{ \epsilon^{2}(\partial_{y}H)^{2}-\hat{M}^{2}\}]
+\frac{T_{{\rm M9}}}{2}H^{2\epsilon/3}\eta_{\mu\nu}\delta(y)=0\ \ 
\label{delgmm}
\\
\frac{\delta S^{{\rm total}}}{\delta \hat{g}_{yy}}&=&
-\frac{1}{4\bar{\kappa}}H^{-7\epsilon/3-1}
\{\epsilon^{2}(\partial_{y}H)^{2}-\hat{M}^{2}\}=0\label{delgyy}
\\
\frac{\delta S^{{\rm total}}}{\delta \hat{g}_{zz}}&=&
-\frac{3}{2\bar{\kappa}}H^{8\epsilon/3}[\epsilon \partial_{y}^{2}H
+\frac{5}{3}H^{-1}
\{\epsilon^{2}(\partial_{y}H)^{2}-\hat{M}^{2}\}]
-\frac{3\cdot T_{{\rm M9}}}{2}H^{8\epsilon/3}\delta(y)
=0.\label{delgzz}
\eeqa
So, 
a solution to these equations is obtained as
\beqa
H=c-\frac{T_{{\rm M9}}\bar{\kappa}}{2\epsilon}
|y|\ (=
c-\frac{\bar{m}}{\epsilon}|y|)
\eeqa
for an arbitrary nonzero $\epsilon$ and a constant $c$.
We note that $H$ is a harmonic function on $y$.
We also note that if one require $H$ to be positive in order to avoid
a singularity at $H=0$,
one must take $c$ to be positive and $\epsilon$ to be
negative.

The remaining target-space field equation 
is that of $M$, which in this case 
is
\beqa
\sqrt{|\hat{g}|}\hat{M}|\hat{k}|^{4}
+\frac{\hat{\epsilon}^{\mu_{1}\cdots\mu_{10}z}}{\sqrt{|\hat{g}|}}
\frac{10}{10!}\partial_{[\mu_{1}}
(i_{\hat{k}}\hat{A}^{(10)})_{\mu_{2}\cdots\mu_{10}]}
=H^{\epsilon-1}\hat{M}-\partial_{y}\hat{A}^{(10)}_{01..8z}=0.
\eeqa
This equation determines (the
field strength of) $\hat{A}^{(10)}$.
It is solved by
\beqa
\hat{A}^{(10)}_{01\cdots 8z}=H^{\epsilon}.
\eeqa
Then, $X^{\mu}$ field equations are satisfied.
Thus, we can obtain 
the M-9-solution with a nontrivial 10-form
by solving the equations of motion with the source terms.
The dimensional reduction of the solution along z gives the D-8-brane
solution\cite{berg3}\cite{bergm9}.

\section{Summary and discussion}
We have constructed the Wess-Zumino action of a single M-9-brane 
wrapped around the compact direction.
We have also obtained 
the M-9-solution with a nontrivial configuration of the 10-form
by solving the equations of motion with the source terms.
This implies the consistency of the M-9-brane action,
not only the WZ term obtained in this paper but also
the scaling factor $|\hat{k}|^{3}$
of the M-9-brane kinetic action 
\cite{bergm9}\cite{eyras1} (to be concrete, see 
(\ref{delgzz})).
Another consistency check can be done 
by considering a ``test'' M-9-brane 
in an M-9-brane background. 
The ``static potential'' $V$ of the test M-9-brane
is obtained by substituting the M-9-brane solution 
obtained in section 4 for the target-space fields of the M-9-brane
effective action, as done in ref.\cite{tsey1}.
Then, putting
the test brane parallel to the background M-9-brane
with an orientation,
we have the potential $V=H^{\epsilon} -H^{\epsilon}=0$.
This implies that the test brane is stable, which is a reasonable result.
So, we conclude that 
the introduction of the 10-form and
the construction of the M-9-brane WZ action is 
achieved in a consistent way.

Furthermore, 
as discussed in 
ref.\cite{hulm9}\cite{meessen1}\cite{eyras1}\cite{sptfilling}%
\cite{exotic},
there are other possibilities of dimensional reduction:
The M-9-brane  
dimensionally reduced along 
the worldvolume direction
but not the isometry direction 
is supposed to give an 8-brane with a gauged direction,
called a ``KK-8A brane\cite{meessen1}\cite{eyras1}.
On the other hand, the M-9-brane
dimensionally reduced along the transverse direction
is to give an 9-brane called an 
``NS-9A brane''\cite{hulm9}\cite{sptfilling}.
The worldvolume actions of the former
is given fully in ref.\cite{exotic} and
that of the latter is presented in
ref.\cite{sptfilling}, but 
both are obtained via dualities.
So, deriving these actions 
via dimensional reduction
from the M-9-brane action, including the WZ
part obtained in this paper,
will reinforce the consistency of the action.
This appears to be possible, 
but we do not discuss this further here.

\parbigskipn

{\large \bf Acknowledgment}

\parbigskipn
I would like to thank Taro Tani for useful discussions
and encouragement.


\appendix

{\large\bf Appendix}

In this appendix we present the relations between the 11D and the 
10D fields via dimensional reduction in the isometry 
direction\cite{berg4}.
(Hats on indices indicate that they are 11-dimensional, and
absence of Hats indicates that they are 10-dimensional.)
The elfbein basis is
\[
\left. 
\begin{array}{cc}
(\hat{e}_{\hat{\mu}}^{\ \hat{a}})=
\left(
\begin{array}{cc}
e^{-\phi/3}e_{\mu}^{\ a} & e^{2\phi/3}C^{(1)}_{\mu}\\
0 & e^{2\phi/3}
\end{array}
\right),
&
(\hat{e}^{\ \hat{\mu}}_{\hat{a}})=
\left(
\begin{array}{cc}
e^{\phi/3}e^{\ \mu}_{a} & -e^{\phi/3}C^{(1)}_{a}\\
0 & e^{-2\phi/3}
\end{array}
\right).
\end{array}
\right.
\]
The 11D metric and the 11D 3-form gauge field are expressed as
\[
\left. 
\begin{array}{cc}
\left\{
\begin{array}{cc}
\hat{g}_{\mu\nu}= &e^{-2\phi/3}g_{\mu\nu}
- e^{4\phi/3}C^{(1)}_{\mu}C^{(1)}_{\nu}, \\
\hat{g}_{\mu z}= & (i_{\hat{k}}\hat{g})_{\mu}=
- e^{4\phi/3}C^{(1)}_{\mu}  \\
\hat{g}_{z z}= &-e^{4\phi/3}
\end{array}
\right. 
&
\left\{
\begin{array}{cc}
\hat{C}_{\mu\nu\rho}= &C^{(3)}_{\mu\nu\rho} \\
\hat{C}_{\mu\nu z}= &(i_{\hat{k}}\hat{C})_{\mu\nu}=B_{\mu\nu}
\end{array}
\right.
\end{array}
\right.
\]
The 11D 6-form gauge field splits as 
\[
\left\{
\begin{array}{cc}
\hat{C}^{(6)}_{\mu_{1}\cdots\mu_{6}}= &
-\tilde{B}^{(6)}_{\mu_{1}\cdots\mu_{6}}, \\
\hat{C}^{(6)}_{\mu_{1}\cdots\mu_{5} z}= &
(i_{\hat{k}}\hat{C}^{(6)})_{\mu_{1}\cdots\mu_{5}}=
C^{(5)}_{\mu_{1}\cdots\mu_{5}}-5C^{(3)}_{[\mu_{1}\mu_{2}\mu_{3}}
B_{\mu_{4}\mu_{5}]},
\end{array}
\right.
\]
where $\tilde{B}^{(6)}$ is the 6-form field dual of $B$.
The 11D 8-form gauge field is considered to give the 10D RR 7-form
\[
(i_{\hat{k}}\hat{N}^{(8)})_{\mu_{1}\cdots\mu_{7}}=
C^{(7)}_{\mu_{1}\cdots\mu_{7}}-7\cdot 5C^{(3)}_{[\mu_{1}\mu_{2}\mu_{3}}
B_{\mu_{4}\mu_{5}}B_{\mu_{6}\mu_{7}]},
\]
and an 8-form.
The 10D IIA 8-form seems to correspond to the dual of the dilaton.
The 11D 2-form gauge parameter $\hat{\chi}_{\mu\nu}$ corresponds to the
RR 2-form gauge transformation parameter, while
and $\hat{\chi}_{\mu z}\equiv\hat{\lambda}_{\mu}$ to $\lambda_{\mu}$.
We note that $\hat{\epsilon}^{\mu_{1}\cdots\mu_{10} z}
=\epsilon^{\mu_{1}\cdots\mu_{10}}$, but 
$\hat{\epsilon}_{\mu_{1}\cdots\mu_{10} z}
=-\epsilon_{\mu_{1}\cdots\mu_{10}}$.
Finally, the 
worldvolume gauge field of the M-9-brane $\hat{b}_{i}$ gives
that of the 10D IIA D-8-brane.

\parbigskipn


\begin{thebibliography}{99}


\bibitem{wit1} E. Witten, 
\np {\bf B443 } (1995) 85, hep-th/9503124.

\bibitem{tow2} P. K. Townsend,
\pl {\bf B350} (1995) 184, hep-th/9501068.

\bibitem{pol2} J. Polchinski,
Phys. Rev. Lett. {\bf 75} (1995) 4724, hep-th/9510017.

\bibitem{rom1} L. Romans, 
\pl {\bf 169B} (1986) 374.

\bibitem{berg3} E. Bergshoeff, M. de Roo, M. B. Green,
G. Papadopoulos and P. K. Townsend, 
\np {\bf B470} (1996) 113, hep-th/9601150.


\bibitem{des1} K. Bautier, S. Deser, M. Henneaux and D. Seminara,
\pl {\bf B406} (1997) 49, hep-th/9704131.

\bibitem{hullalg} C. M. Hull,
Nucl. Phys. {\bf B509} (1998) 216, hep-th/9705162.

\bibitem{towalg} P. K. Townsend,
Cargese Lectures 1997,
hep-th/9712004.

\bibitem{hlw1} P. S. Howe, N. D. Lambert and P. C. West,
Phys. Lett. {\bf B416} (1998) 303, hep-th/9707139.
 
\bibitem{berg4} E. Bergshoeff, Y. Lozano and T. Ortin,
\np {\bf B518} (1998) 363, hep-th/9712115.


\bibitem{loz1} E. Bergshoeff, E. Eyras and Y. Lozano,
\pl {\bf B430} (1998) 77, hep-th/9802199.

\bibitem{hull3} C. M. Hull,
JHEP {\bf 9811} (1998) 027, hep-th/9811021.


\bibitem{bergm9} E. Bergshoeff and J. P. van der Schaar,
Class. Quant. Grav. {\bf 16} (1999) 23, hep-th/9806069.

\bibitem{pol1} J. Polchinski and E. Witten,
\np {\bf B460} (1996) 525, hep-th/9510169.



\bibitem{eyras1} E. Eyras and Y. Lozano,
``{\it Brane Actions and String Dualities}'', 
hep-th/9812225.

\bibitem{exotic} E. Eyras and Y. Lozano,
``{\it Exotic Branes and Nonperturbative Seven Branes}'', 
hep-th/9908094.

\bibitem{sptfilling} E. Bergshoeff, E. Eyras, R. Halbersma,
C. M. Hull, Y. Lozano and J. P. van der Schaar,
``{\it Spacetime-Filling Branes and Strings with 
Sixteen Supercharges}'', 
hep-th/9812224, to appear in Nucl. Phys. B.


\bibitem{lozbrantiba} L. Houart and Y. Lozano,
``{\it Type II Branes from Brane-Antibrane in M-theory}'',
hep-th/9910266.


\bibitem{bergmassiveT} E. Bergshoeff and M. de Roo,
Phys. Lett. {\bf B380} (1996) 265, hep-th/9603123.


\bibitem{DWZ} M. B. Green, C. M. Hull and P. K. Townsend,
Phys. Lett. {\bf B382} (1996) 65, hep-th/9604119.

\bibitem{superDbr} E. Bergshoeff and P. K. Townsend, 
Nucl. Phys. {\bf B490} (1997) 145, hep-th/9611173.

\bibitem{topomassive} E. Bergshoeff, P. M. Cowdall and 
P. K. Townsend, 
\pl {\bf B410} (1997) 13, hep-th/9706094.

\bibitem{kaluzakleinm} E. Bergshoeff, B. Janssen and T. Ortin,
Phys. Lett. {\bf B410} (1997) 131, hep-th/9706117.

\bibitem{tsey1} A. A. Tseytlin, \np {\bf 487} (1997) 141, 
hep-th/9609212.

\bibitem{hulm9} C. M. Hull, JHEP {\bf 9807} (1998) 018, 
hep-th/9712075.

\bibitem{meessen1} P. Meessen and T. Ortin,
\np {\bf 541} (1999) 195, hep-th/9806120.

\end{thebibliography}
\end {document}